\documentclass[epj]{webofc}
\usepackage[varg]{txfonts}   
\usepackage{amsmath}
\usepackage{graphicx,color}
\woctitle{International Conference on New Frontiers in Physics 2013}
\begin{document}
\title{Interplay between chromoelectric and chromomagnetic gluons
in Yang-Mills thermodynamics}
%
%

\author{Chihiro Sasaki\inst{1}\fnsep\thanks{
\email{sasaki@fias.uni-frankfurt.de}} 
}

\institute{
Frankfurt Institute for Advanced Studies,
D-60438 Frankfurt am Main,
Germany
          }

\abstract{%
We propose an effective theory of SU(3) gluonic matter
where interactions between color-electric and color-magnetic gluons
are constrained by the center and scale symmetries.
Through matching to the dimensionally-reduced magnetic theories,
the magnetic gluon condensate qualitatively changes its thermal behavior 
above the critical temperature. 
We argue its phenomenological consequences for the thermodynamics,
in particular the dynamical breaking of scale invariance.
}
\maketitle
%

\section{Magnetic confinement}
\label{mag}

In Yang-Mills (YM) theories at finite temperature $T$ 
the magnetic screening mass is dynamically generated as a ultra-soft
scale $g^2T$~\cite{Linde,GPY}. Consequently, the magnetic sector
remains non-trivial in the high temperature phase
and in fact, the spatial string tension is non-vanishing for all 
temperatures~\cite{sst,sst2}, indicating certain confining properties.
Due to this residual interaction, equations of state (EoS) deviate 
from their Stefan-Boltzmann limit at high temperature. 
The interaction measure $I(T)/T^2T_c^2$ obtained from SU(3) lattice 
simulations is nearly constant above the deconfinement critical 
temperature $T_c$, in the range $T_c < T < 5\,T_c$~\cite{lat,latYM}.

Several scenarios have been proposed to explain this non-perturbative 
nature, see e.g.~\cite{dim2,CMS,Carter,fuzzy,matrix}. The major assumption
in those models is an effective gluon mass being {\it constant} in the
intermediate $T$ range. In \cite{SMR} we propose a different effective
theory where the non-trivial behavior of $I(T)/T^2$ emerges {\it dynamically}
from chromomagnetic gluons, without relying on
any assumption for the gluon thermal mass.

Using the three-dimensional YM theories~\cite{AP,Nadkarni,Landsman,Kajantie},
thermal behavior of the magnetic gluon condensate at high temperature 
is found as~\cite{Agasian}
\begin{equation}
\langle H \rangle = c_H \left(g^2(T)T\right)^4\,,
\end{equation}
with
\begin{equation}
c_H = \frac{6}{\pi}c_\sigma^2 c_m^2\,.
\end{equation}
The constants $c_\sigma$ and $c_m$ appear in the spatial string tension 
and in the magnetic gluon mass as
\begin{equation}
\sqrt{\sigma_s(T)} = c_\sigma g^2(T)T\,,
\quad
m_g(T) = c_m g^2(T)T\,.
\end{equation}
For $SU(3)$ YM theory $c_\sigma = 0.566$~\cite{lat} and
$c_m = 0.491$~\cite{3dYM}.

The non-vanishing string tension $\sigma_s$ may support to consider
a system where hadronic states, glueballs, can survive in
deconfined phase. The scalar glueballs can be introduced as dilatons
associated with the scale symmetry breaking 
through the potential~\cite{schechter}
\begin{equation}
V_\chi
= \frac{B}{4}\left(\frac{\chi}{\chi_0}\right)^4
\left[ \ln\left(\frac{\chi}{\chi_0}\right)^4 - 1 \right]\,,
\end{equation}
where $B$ is the bag constant and $\chi_0$ is a dimensionful constant.
The two parameters, $B$ and $\chi_0$, are fixed to reproduce the vacuum energy
density ${\mathcal E} = \frac{1}{4}B = 0.6$ GeV fm$^{-3}$
and the vacuum glueball mass $M_\chi = 1.7$ GeV~\cite{narison,sexton},
leading to $B = (0.368\,\mbox{GeV})^4$ and $\chi_0 = 0.16\,\mbox{GeV}$.

The confinement-deconfinement phase transition is characterized by
the Polyakov loop $\Phi$,
\begin{equation}
\Phi
=
\frac{1}{N_c}\mbox{tr}\hat{L}\,,
\quad
\hat{L}
=
{\mathcal P}\exp\left[i\int_0^{1/T}d\tau A_4(\tau,\vec{x})\right]\,,
\end{equation}
which is is an order parameter of dynamical 
breaking of $Z(N_c)$ symmetry~\cite{mclerran}.
The two scalar fields, $\chi$ and $\Phi$, are mixed and their potential
should be manifestly invariant under $Z(N_c)$ and scale transformation.
For $N_c=3$, the most general form is given as~\cite{Sannino},
\begin{equation}
V_{\rm mix} = \chi^4\left(
G_1\bar{\Phi}\Phi + G_2\left(\bar{\Phi}^3+\Phi^3\right)
{}+ G_3\left(\bar{\Phi}\Phi\right)^2 + \cdots
\right)\,,
\label{mix}
\end{equation}
with unknown coefficients $G_i$. In the following calculation, 
we take only the first term.

\section{Trace anomaly}
\label{int}

We formulate the model accounting for
a mixing between chromoelectric and chromomagnetic gluons as
\begin{equation}
\Omega = \Omega_g + \Omega_\Phi
{}+ V_\chi + V_{\rm mix} + c_0\,.
\end{equation}
The color-electric gluon part $\Omega_g$ is given
in the presence of a uniform field $A_0$ as
\begin{equation}
\Omega_g
= 2T \int\frac{d^3p}{(2\pi)^3}\ln
\left( 1 + \sum_{n=1}^8C_n\, e^{-np/T}
\right)\,,
\end{equation}
where the coefficients $C_n$ are certain functions of the $SU(3)$
group characters~\cite{SR}.
The Haar measure part is
\begin{equation}
\Omega_\Phi
=
-a_0T\ln\left[ 1 - 6\bar{\Phi}\Phi + 4\left( \Phi^3 + \bar{\Phi}^3\right)
{}- 3\left(\bar{\Phi}\Phi\right)^2\right]\,.
\end{equation}
We take the following mixing term,
\begin{equation}
V_{\rm mix}
= G \left(\frac{\chi}{\chi_0}\right)^4 \bar{\Phi}\Phi\,.
\end{equation}
Requiring that a first-order phase transition appears at $T_c=270$ MeV
as found in the lattice results~\cite{lat},
one finds
$a_0 = (0.184\,\mbox{GeV})^3$, $c_0 = (0.244\,\mbox{GeV})^4$
and $G = (0.206\,\mbox{GeV})^4$.

At high temperature, the theory in four dimensions should match the 
three-dimensional effective theory via dimensional reduction.
We postulate the matching condition for the gluon condensate as
\begin{equation}
\frac{\langle\chi\rangle}{\chi_0}
= \left(\frac{\langle H \rangle}{H_0}\right)^{1/4}\,.
\end{equation}
We use the two-loop running coupling,
\begin{equation}
g^{-2}(T)
=
2b_0\ln\frac{T}{\Lambda_\sigma}
{}+ \frac{b_1}{b_0}\ln\left(2\ln\frac{T}{\Lambda_\sigma}\right)\,,
\quad
b_0
=
\frac{11}{16\pi^2}\,,
\quad
b_1 = \frac{51}{128\pi^2}\,,
\end{equation}
with $\Lambda_\sigma = 0.104\,T_c$~\cite{lat}.
In this model, the  changeover  in  the temperature dependence of the
magnetic condensate  appears by construction.
In YM theories, however, such behavior should emerge dynamically.
A qualitative change of the thermal gluon can be in fact
seen in the spatial string tension at  $T \sim 2\,T_c$~\cite{Agasian}.

This also affects the EoS.
One finds an additional contribution
to the interaction measure from $\langle H\rangle$  as
\begin{equation}
\delta I
= -B\frac{\langle H\rangle}{H_0}
{}+ \left(2b_0 + \frac{b_1}{b_0}
\frac{1}{\ln\left(T/\Lambda_\sigma\right)}\right)
\frac{\langle H\rangle}{g^4(T)H_0}\,.
\label{delta}
\end{equation}
The $I/T^2T_c^2$ is monotonically decreasing even at high temperature when
no matching to the 3-dim YM is made.
The magnetic contribution generates a $T^2$ dependence and
this results in a plateau-like behavior emerging in  $I/T^2T_c^2$
at moderate temperature, $T/T_c \sim 2$-$4$.
This property appears due to the residual chromomagnetic interaction 
encoded in the dilaton, $\chi^4 \sim H$.
The obtained behavior of $I/T^2T_c^2$ with temperature
qualitatively agrees
with the latest lattice data~\cite{latYM}.

\section{Summary}
\label{sum}

We have presented an effective theory of SU(3) YM thermodynamics 
implementing the major global symmetries, the center and scale symmetries.
This naturally allows a mixing between the
Polyakov loop and the dilaton field. Consequently, the magnetic confinement is
effectively embedded and results in deviations of the EoS from their
Stefan-Boltzmann limit at high temperature.

Matching to the 3-dimensional YM theory suggests
the gluon condensate  increasing with temperature in deconfined phase.
We have illustrated that this changeover  
becomes transparent
in the interaction measure $I={\mathcal E}-3P$ normalized by $T^2T_c^2$,
rather than by $T^4$.
We have found that the role of the magnetic gluon is alternative to 
the Hard Thermal Loop contribution.

\section*{Acknowledgments}

C.~S. acknowledges partial support by the Hessian
LOEWE initiative through the Helmholtz International
Center for FAIR (HIC for FAIR).

\end{document}